\documentclass[a4paper, amsfonts, amssymb, amsmath, reprint, showkeys, twoside,superscriptaddress, onecolumn, nofootinbib]{revtex4-2}
\usepackage[colorlinks=true, allcolors=blue]{hyperref}
\usepackage[capitalize,nameinlink]{cleveref}
\usepackage{orcidlink}
\usepackage{latexsym}
\usepackage{titlesec}
\usepackage{amsmath}
\usepackage{amsfonts}
\usepackage{amssymb}
\usepackage{amsthm}
\usepackage{physics}
\usepackage{enumitem}
\usepackage{caption}
\usepackage{tikz}
\usetikzlibrary{quantikz2}

\usepackage{graphicx}
\usepackage{dcolumn}
\usepackage{caption}
\usepackage{subcaption}
\usepackage{bm}
\usepackage{algorithm}
\usepackage{algpseudocode}
\usepackage{xcolor}
\usepackage{caption}
\usepackage{siunitx}
\usepackage{multirow}
\usepackage{algorithm}
\usepackage{algpseudocode}
\usepackage{mdframed}
\floatname{algorithm}{}

\algtext*{EndIf}
\algtext*{EndFor}
\algtext*{EndWhile}
\algtext*{EndProcedure}
\algtext*{EndFunction}

\AtBeginDocument{\usepackage{booktabs}} 

\definecolor{XanaduBlue}{HTML}{4D53C8}
\definecolor{XanaduRed}{HTML}{D7333B}
\definecolor{XanaduGreen}{HTML}{52BD7C}
\definecolor{XanaduOrange}{HTML}{E2A300}
\definecolor{XanaduPurple}{HTML}{8C4ACF}

\def\be{\begin{equation}} %
\def\ee{\end{equation}} %
\def\bea{\begin{eqnarray}} %
\def\eea{\end{eqnarray}} %
\begin{document}
\author{Robert A. Lang \orcidlink{0000-0002-4345-3566}}
\email{robert.lang@xanadu.ai}
\affiliation{Xanadu, Toronto, ON, M5G 2C8, Canada}

\author{Paarth Jain \orcidlink{0009-0005-9107-4789}}
\affiliation{Xanadu, Toronto, ON, M5G 2C8, Canada}
\affiliation{Chemical Physics Theory Group, Department of Chemistry,
  University of Toronto, Toronto, Ontario, M5S 3H6, Canada}

\author{Juan Miguel Arrazola \orcidlink{0000-0002-0619-9650}}
\affiliation{Xanadu, Toronto, ON, M5G 2C8, Canada}

\author{Danial Motlagh \orcidlink{0009-0003-7655-4341}}
\email{danial.motlagh@xanadu.ai}
\affiliation{Xanadu, Toronto, ON, M5G 2C8, Canada}

\title{Quantum algorithm for simulating non-adiabatic dynamics at metallic surfaces}

\begin{abstract}
Non-adiabatic dynamics at molecule-metal interfaces govern diverse and technologically important phenomena, from heterogeneous catalysis to dye-sensitized solar energy conversion and charge transport across molecular junctions. Realistic modeling of such dynamics necessitates taking into account various charge and energy transfer channels involving the coupling of nuclear motion with a very large number of electronic states, leading to prohibitive cost using classical computational methods. In this work we introduce a generalization of the Anderson–Newns Hamiltonian and develop a highly optimized quantum algorithm for simulating the non-adiabatic dynamics of realistic molecule-metal interfaces. Using the PennyLane software platform, we perform resource estimations of our algorithm, showing its remarkably low implementation cost for model systems representative of various scientifically and industrially relevant molecule-metal systems.  Specifically, we find that time evolution for models including $100$ metal orbitals, $8$ molecular orbitals, and $20$ nuclear degrees of freedom, requires only $271$ qubits and $7.9 \times 10^7$  Toffoli gates for $1000$ Trotter steps, suggesting non-adiabatic molecule-metal dynamics as a fruitful application of first-generation fault-tolerant quantum computers. 
\end{abstract}

\maketitle

\section{Introduction}

The current landscape of quantum applications presents a fundamental challenge. High-impact proposals, such as large-scale electronic structure calculations \cite{goings2022reliably, zini2023quantum, nguyen2024quantum, rubin2024quantum, watts2024fullerene, bartschi2024potential}, face a long and costly path to realization due to their large resource requirements, while proposals for applications with a shorter horizon, such as simulation of spin models~\cite{childs2018toward, dalzell2023quantum, kan2024resource, rosenberg2024dynamics}, are not generally believed to result in transformative impact. Quantum dynamics stands out as a domain where the tradeoff between resource requirements and impact becomes far more favorable than for fixed-point energy calculations. Classically, simulating real-time quantum dynamics is significantly harder than static problems, widening the gap between classical and quantum methods, and enabling quantum advantage at smaller system sizes. This translates directly to lower resource demands for meaningful applications. At the same time, these simulations remain highly impactful as many key functional properties of materials, such as charge and energy transfer, are inherently dynamical. Accurate quantum simulations of these processes can provide predictive insights that bypass experimental bottlenecks and accelerate materials discovery.\\

A central and especially demanding class of dynamical problems arises in the non-adiabatic regime, when the motion of electrons can no longer be treated as instantaneously adjusting to nuclear positions, resulting in coupled electron–nuclear dynamics and transitions between electronic states during molecular motion. One domain of high industrial relevance where non-adiabatic dynamics is ubiquitous is in molecular photodynamics, where predictive modeling could accelerate the discovery of high-performance optical materials for various technologies \cite{motlagh2025quantum}. Another equally impactful and more challenging setting where non-adiabatic effects are intrinsic is the interaction of molecular species with metallic surfaces, where nuclear motion can entangle with a quasi-continuous manifold of electronic states, requiring a description of nuclei evolving on many potential energy surfaces, invalidating the Born–Oppenheimer approximation \cite{wodtke2004electronically, muino2013dynamics, wodtke2016electronically, dou2020nonadiabatic}. These interfacial processes are central to a broad range of industrial and scientific applications, such as heterogeneous catalysis \cite{luo2016electron, dorenkamp2018hydrogen}, electrochemical energy storage \cite{lam2020theory, lee2022hot}, molecular junctions for nanoelectronics \cite{joachim2005molecular, erpenbeck2015effect, kaspar2022nonadiabatic}, dye-sensitized solar cells \cite{kondov2007quantum, duncan2007theoretical, akimov2013theoretical}, and corrosion inhibition \cite{katz1999non, behler2008nonadiabatic, shimakura2025quantum}.\\

Despite its potential impact, the characterization of non-adiabatic dynamics at metallic surfaces has proved challenging both experimentally \cite{d1983reactive, park2019kinetics, borodin2022quantum}  and computationally \cite{dou2020nonadiabatic, gardner2023assessing}. A plethora of classical simulation methods have been proposed; these include mixed quantum–classical methods like electronic friction models \cite{head1995molecular, dou2017born}, Ehrenfest dynamics \cite{grotemeyer2014electronic, tully2023ehrenfest}, or surface hopping approaches \cite{shenvi2009dynamical, shenvi2009nonadiabatic, miao2019comparison}, as well as fully quantum methods such as those based on the hierarchical equations of motion \cite{tanimura2020numerically, bai2024hierarchical, xu2019non, dan2023theoretical, ke2023current, zhang2025multiset}, or multilayer multiconfigurational time-dependent Hartree (ML-MCTDH)~\cite{beck2000multiconfiguration, wang2003multilayer, sasmal2020non}. However, scalability of these methods comes at the cost of reduced accuracy; relying on simplifying assumptions, regime-specific approximations, or severely reducing the dimensionality of the problem which omit key aspects of the underlying physics. These limitations suggest that quantum computers could provide a transformative means of accurately describing complex molecular processes at metal surfaces.\\

In this work we introduce a generalization of the Anderson-Newns Hamiltonian, the standardly employed model for non-adiabatic dynamics of molecules at metal surfaces, by including an arbitrary number of molecular electronic states, nuclear degrees of freedom, and electron-electron correlation effects to more accurately capture the underlying physics. We then develop a highly optimized Trotter-based quantum algorithm for implementing the time evolution under the generalized Anderson-Newns (GAN) Hamiltonian. The algorithm inherits its favorable resource requirements from our previously developed algorithm for vibronic dynamics \cite{motlagh2025quantum} and several novel optimizations tailored specifically to the GAN model. Lastly, we provide resource estimates for models of various complexity in \cref{tab:resource_estimation}. These resource estimates were computed using the PennyLane software platform, which supports all underlying building blocks of the algorithm, allowing for rapid and accurate estimates. We find that a model including $100$ metal orbitals, $8$ molecular orbitals, and $20$ nuclear degrees of freedom, requires only $271$ qubits and $7.85\times 10^7$ Toffoli gates for $1000$ Trotter steps. It is worth noting that our resource estimates do not take into account the sparsity present in real-world systems which can further reduce the costs presented.\\

The rest of the paper is structured as follows. We introduce the GAN Hamiltonian in \cref{sec:theory}, and describe the quantum algorithm for its time evolution in \cref{sec:algorithm}. We then discuss the preparation of initial states and the extraction of observables in \cref{sec:i/o}. We outline various technlogically relevant potential applications and present resource estimates for models of various complexity in \cref{sec:applications}. The paper is concluded with discussions and future outlook in \cref{sec:discussion}. 

\begin{figure}[t] 
    \centering
    \includegraphics[width=0.925\linewidth]{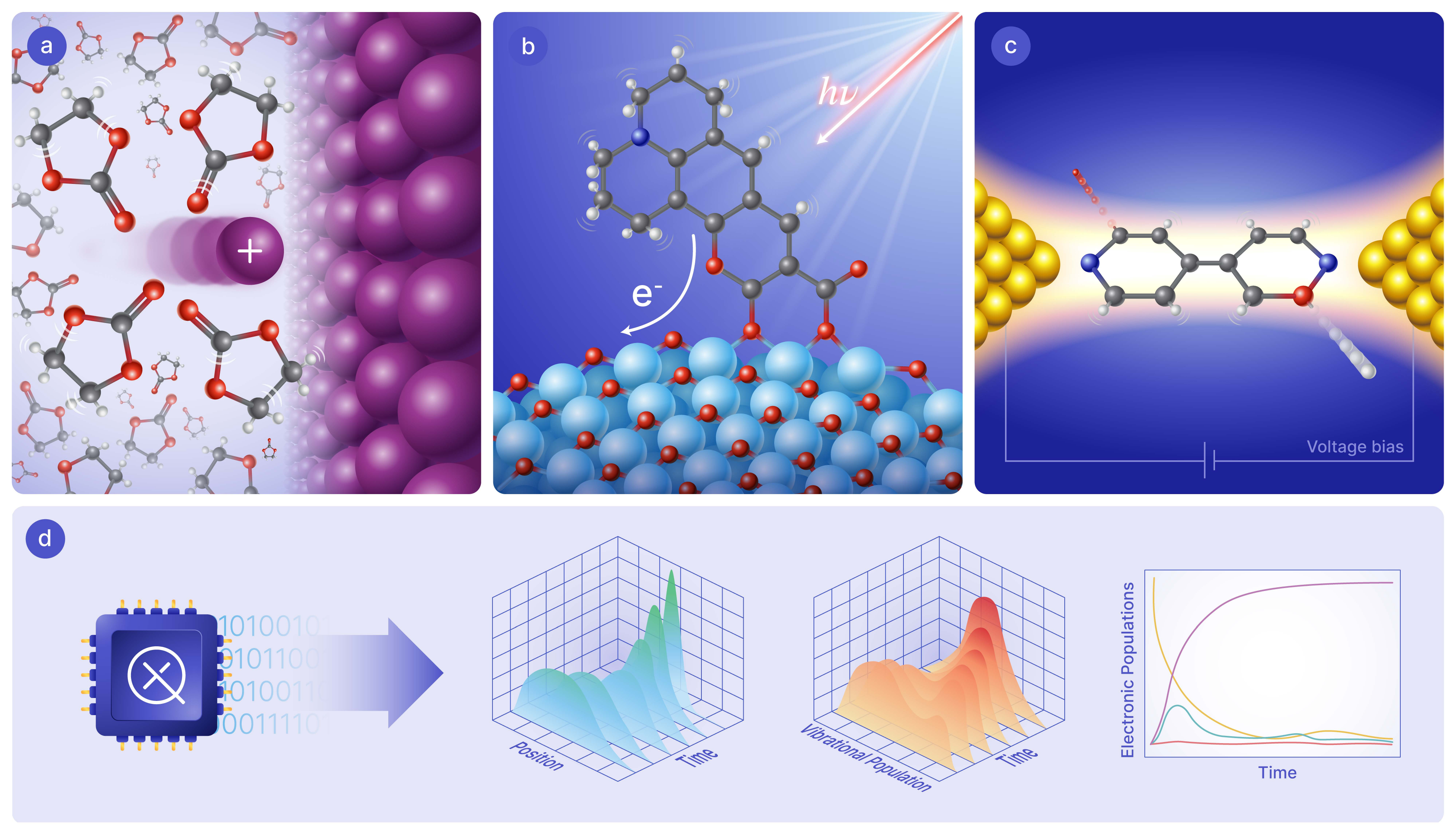}
    \caption{Graphical depiction of various simulation problems that can be tackled by our new algorithm and examples of relevant observables. In the top panels, representative systems involving the non-adiabatic dynamics of molecules interacting with interfaces are depicted: a) coupled ion-electron transfer at a battery electrode in the presence of solvent, b) photo-induced charge transfer from a chromophore to a semiconductor nanoparticle in a dye-sensitized solar cell, and c) current-induced bond rupture in a molecular junction under voltage bias. The bottom panel d) shows examples of time-resolved nuclear and electronic observables of interest extracted via quantum simulation.}
    \label{fig:hero}
\end{figure}

\newpage
\section{Theory} \label{sec:theory}
\subsection{Generalized Anderson-Newns Hamiltonian}\label{app:general_model}

The Anderson–Newns (AN) Hamiltonian serves as the standard model for describing non-adiabatic dynamics at metal–molecule interfaces \cite{newns1969self, anderson1961localized, dou2020nonadiabatic}. As conventionally employed, the AN Hamiltonian describes a single electronic level of a molecule, e.g., within a valence orbital, and one or two nuclear degrees of freedom coupled to the electronic continuum of a metal. However, processes at metal–molecule interfaces can involve several electronic orbitals, electron-electron correlation effects, and non\mbox{-}negligible contributions from multiple nuclear degrees of freedom (DOFs) \cite{baldea2012extending, ke2023current}. To capture a more accurate description of the non-adiabatic dynamics at metallic surfaces, we consider a multi-orbital and multi-mode generalized Anderson–Newns (GAN) Hamiltonian
\begin{align} \label{eq:h_gan}
\hat H = \hat H_{\rm{mol}} + \hat H_{\rm{metal}} + \hat H_{\rm{int}},
\end{align}
where $\hat H_{\rm{mol}}$ is the molecular Hamiltonian, 
\begin{align} \label{eq:h_mol}
\hat H_{\rm{mol}} & = \sum_{\kappa=0}^{M-1} \frac{\hat P_{\kappa}^2}{2m_{\kappa}} 
 + \sum_{\substack{i,j \in \mathcal{M}}} \hat U_{ij}(\boldsymbol{\vec Q}) \hat a_i^\dagger \hat a_j + \sum_{i,j \in \mathcal{M}} \hat V_{ij} (\boldsymbol{\vec Q}) \hat a_i^\dagger \hat a_i \hat a_j^\dagger \hat a_j\, + \,\hat U_0 (\boldsymbol{\vec{Q}}),
\end{align}
where $\hat P_{\kappa}$ is the momentum operator for the $\kappa^{\rm{th}}$ nuclear coordinate with mass $m_{\kappa}$, $\boldsymbol{\vec Q} = (\hat Q_0,\hat Q_1,...,\hat Q_{M-1})$ denotes the position operators of all nuclear coordinates, and $\hat U_0(\boldsymbol{\vec{Q}})$ is a reference nuclear potential. $\hat U_{ij}(\boldsymbol{\vec Q})$ and $\hat V_{ij}(\boldsymbol{\vec Q})$ describe, respectively, the electronic couplings and the electron–electron repulsion within the molecule as functions of the nuclear coordinates. The specific form of these operators depends on the specific problem instance, however, we discuss a few frequently employed forms of these operators in \cref{sec:nuc_potentials}. $\mathcal{M}$ denotes the set of electronic basis indices of the molecule, and $\hat a_i^\dagger$($\hat a_i$) are the fermionic creation(annihilation) operators for the $i^{\rm{th}}$ basis state. Depending on the modeling context, $\hat a_i^\dagger$ and $\hat a_i$ in \cref{eq:h_mol} are interpreted either as acting on one-electron spin-orbitals (e.g., of localized frontier orbitals), or as acting on effective fermionic modes that encode particle-number--changing transitions between many-body diabatic states (i.e., neutral and ionic configurations). In both representations, the number of molecular fermionic modes is typically small, as the non-adiabatic charge-transfer dynamics at molecule–metal interfaces is dominated by a limited subset of active electronic degrees of freedom, such as orbitals in energetic proximity to the metal Fermi level, or a small number of low-lying charge configurations.\\ 

Next, the metal system is described by
\begin{align} \label{h_metal}
\hat H_{\rm{metal}} = \sum_{i \in \mathcal{B}} \epsilon_{i} \hat a_{i}^\dagger \hat a_{i}, 
\end{align}
where $\mathcal{B}$ denotes the set of metallic orbital indices with energies $\epsilon_i$. The molecule-metal interaction is given by 
\begin{align} \label{eq:h_mol_metal}
\hat H_{\rm{int}} = \sum_{i \in \mathcal{M}} \sum_{j \in \mathcal{B}} \hat W_{ij} (\boldsymbol{\vec Q}) \left( \hat a_i^\dagger \hat a_j + \hat a_j^\dagger \hat a_i \right),
\end{align}
where the coupling $\hat W_{ij}(\boldsymbol{\vec Q})$ modulates the molecule-metal electron transfer as a function of the nuclear coordinates.

\subsection{Frequently employed potentials}\label{sec:nuc_potentials}

Having introduced the GAN Hamiltonian in its full generality, here we discuss a few frequently employed forms of the nuclear-dependent functions $\hat U_0(\boldsymbol{\vec{Q}})$, $\hat U_{ij}(\boldsymbol{\vec Q})$, $\hat V_{ij}(\boldsymbol{\vec Q})$, and $\hat W_{ij}(\boldsymbol{\vec Q})$. For this work, we consider $M-1$ internal nuclear DOFs for the molecule, $\hat Q_0, \hat Q_1,..., \hat Q_{M-2}$, and one translational DOF, $\hat Q_{M-1}$, which parameterizes the molecule-metal distance. The internal coordinates can be further grouped into $M'$ \textit{non-reactive} modes, $\hat Q_0, \hat Q_1,..., \hat Q_{M'-1}$, and $M-M'-1$ \textit{reactive} modes, $\hat Q_{M'}, \hat Q_{M'+1},..., \hat Q_{M-2}$, where reactive modes are expected to undergo large displacements and be directly involved in significant chemical changes, such as high amplitude motion along stretching modes, whereas non-reactive modes do not directly define dynamical paths, but still have essential involvement in relaxation channels and mediating vibronic coupling.\\

Non-reactive modes are well described by a Taylor expansion around the equilibrium structure, while for reactive DOFs which experience highly anharmonic potentials, it is common to employ physically motivated analytical potential forms giving us

\begin{align} \label{eq:U}
\hat U_{ij}(\boldsymbol{\vec Q})
&=
\begin{cases}
\displaystyle \lambda^{(i,j)} + \sum_{\kappa = 0}^{M-1} \,a_{\kappa}^{(i,j)} \hat Q_{\kappa} + \sum_{\kappa, \kappa{'} = 0}^{M-1} \,b_{\kappa, \kappa{'}}^{(i,j)} \hat Q_{\kappa} \hat Q_{\kappa{'}}\, + \,\cdots\,, & i\neq j\,\,\\[15pt]
\displaystyle \lambda^{(i)} + \sum_{\kappa = M'}^{M-1} f_{\kappa}^{(i)}(\hat Q_{\kappa}) +\sum_{\kappa = 0}^{M'-1} \,a_{\kappa}^{(i)} \hat Q_{\kappa}  + \sum_{\kappa, \kappa{'} = 0}^{M-1} \,b_{\kappa, \kappa{'}}^{(i)} \hat Q_{\kappa} \hat Q_{\kappa{'}}\, + \,\cdots\,, & i=j\,\,,
\end{cases}
\end{align}
where $f_{\kappa}^{(i)}(\hat Q_{\kappa})$ are the physically motivated analytical potentials. While these functions can in principle be expanded as polynomials, we find that for practical instances, it is more efficient to implement them directly as discussed later in \cref{sec:algorithm}. An example of a frequently used $f_{\kappa}^{(i)}(\hat Q_{\kappa})$ for bond-dissociative modes is the Morse potential
\begin{align} \label{eq:morse_potential}
f_{\kappa}^{(i)}(\hat Q_{\kappa}) =  D_{\kappa}^{(i)} \left( 1 - e^{-A_{\kappa}^{(i)}(\hat Q_\kappa - r_{\kappa}^{(i)})}  \right ) ^2,
\end{align}
where $D_{\kappa}^{(i)}$ is the dissociation energy, $A_{\kappa}^{(i)}$ is the bond stiffness, and $r_{\kappa}^{(i)}$ is the equilibrium position. Alternatively, potentials could be repulsive-type exponential functions,
\begin{align} \label{eq:exp_potential}
f_{\kappa}^{(i)}(\hat Q_{\kappa}) = D_{\kappa}^{(i)} e^{-A^{(i)}_{\kappa}(\hat Q_{\kappa} - r_{\kappa}^{(i)})},
\end{align}
where $D_\kappa^{(i)}$, $A_{\kappa}^{(i)}$, and $r_{\kappa}^{(i)}$ are respectively a scaling factor, a decay parameter, and a reference position. We assume the reference potential $\hat U_{0}(\boldsymbol{\vec{Q}})$ takes the same general form as the $i=j$ case of \cref{eq:U}. Next, $\hat V_{ij}(\boldsymbol{\vec Q})$ can also be written as a Taylor expansion of nuclear DOF of the molecule
\begin{equation} \label{eq:V}
    \hat V_{ij}(\boldsymbol{\vec Q}) = h^{(ij)} + \sum_{\kappa = 0}^{M-1} \,c_{\kappa}^{(ij)} \hat Q_{\kappa} + \sum_{\kappa, \kappa{'} = 0}^{M-1} \,d_{\kappa, \kappa{'}}^{(ij)} \hat Q_{\kappa} \hat Q_{\kappa{'}}\, + \,\cdots\,.
\end{equation}
Lastly, $\hat W_{ij}(\boldsymbol{\vec Q})$ is typically written only in terms of the translation coordinate $Q_{M-1}$, as the strength of the molecule-metal interaction most strongly depends on their distance. A commonly employed form is
\begin{align} \label{eq:W}
\hat W_{ij}(\boldsymbol{\vec Q})=w_{ij} \left( \frac{(1-a)}{2} \left( 1 - \tanh\left(\frac{\hat Q_{M-1} - \tilde{r}}{b}\right) \right) + a\right),
\end{align}
where $\{w_{ij}, a, b, \tilde{r}  \}$ are constants. However, note that in general $\hat W_{ij}(\boldsymbol{\vec{Q}})$ could be multivariate.

\subsection{Discretization and encoding}\label{sec:discretization}

To represent states of the molecule-metal system on a quantum computer, we consider the total Hilbert space of the electronic Fock spaces of the molecule and the metal, and the Hilbert space of $M$ nuclear degrees of freedom. The Fock spaces are encoded to qubit registers using the standard Jordan-Wigner encoding \cite{jordan1928paulische}. The nuclear Hilbert space is the $M$-fold tensor product of the individual nuclear DOF spaces, which are encoded using a discretized real-space grid representation \cite{macridin2018digital, motlagh2025quantum}. Concretely, the position operator for each mode is discretized using $K = 2^{k}$ gridpoints with grid spacing $\Delta = \sqrt{2 \pi / K}$, and the eigenvectors correspond to $k$-qubit computational basis states 
\begin{align} \label{eq:q_coordinate}
\hat Q_{\kappa} \ket{x}_{\kappa} = \Delta \cdot x\, \ket{x}_{\kappa},
\end{align}
for integers $x \in \{-\frac{K}{2}, -\frac{K}{2}+1, \hdots, \frac{K}{2}-1  \}$, encoded in the computational basis state $\ket{x}_{\kappa}$ using a two's complemented signed integer representation. Therefore, the total number of qubits required to encode the full molecule-metal system is $N_{\rm{metal}} + N_{\rm{mol}} + M\cdot k$. In practice, $k$ can be different for different nuclear DOFs, with translational and reactive modes generally requiring larger grids than non-reactive modes, and our approach trivially handles the case of non-uniform grid sizes. Lastly, the momentum operator $\hat P_{\kappa}$ is related to $\hat Q_{\kappa}$ via the quantum Fourier transform (QFT)
\begin{equation}\label{eq:qft_diagonalization}
    \hat P_{\kappa} = \text{QFT}^\dagger\,\, \hat Q_{\kappa} \,\,\text{QFT}.
\end{equation}

\newpage
\section{Quantum algorithm} \label{sec:algorithm}

In this section we describe our Trotter-based algorithm for performing time-evolution under the GAN Hamiltonian from \cref{eq:h_gan}. At a high-level, the algorithm works by partitioning the potential term into commuting groups based on the commutation relation of the fermionic creation and annihilation operators. Each group is then diagonalized using a Clifford circuit, and subsequently exponentiated using arithmetic and phase gradient operations. We further introduce an array of optimizations to significantly reduce the implementation cost. These include the use of caching techniques to compute each nuclear-dependent function only once per Trotter step, independent of the number of fragments, and the introduction of coefficient accumulation circuits that allow the simultaneous exponentiation of all terms in a commuting group using a single phase gradient operation for each function. Our optimizations culminate in remarkably low simulation overheads for systems of practical relevance as shown in \cref{sec:applications}. We now lay out our quantum algorithm in more details.\\

Given a system with $N_{\text{mol}} = |\mathcal{M}|$ electronic basis states for the molecule, and $N_{\text{metal}} = |\mathcal{B}|$ metallic orbitals, we re-write the GAN Hamiltonian as
\begin{align}
\hat H = \sum_{\kappa=0}^{M-1} \frac{\hat P_{\kappa}^2}{2m_{\kappa}} \,+\, \sum_{i \in \mathcal{B}} \epsilon_{i} \hat a_{i}^\dagger \hat a_{i} 
 + \sum_{\substack{i \in \mathcal{M} \\ j \in \mathcal{M} \cup \mathcal{B}}}
 \hat g_{ij}(\boldsymbol{\vec Q}) \left( \hat a_i^\dagger \hat a_j + \hat a_j^\dagger \hat a_i  \right)+ \sum_{i,j \in \mathcal{M}} \hat V_{ij} (\boldsymbol{\vec Q}) \hat a_i^\dagger \hat a_i \hat a_j^\dagger \hat a_j, \,+\,\, \hat U_0 (\boldsymbol{\vec{Q}}),
\end{align}
where we have defined
\begin{align}
\hat g_{ij}(\boldsymbol{\vec Q}) &=
    \begin{cases}
    \hat U_{ij}(\boldsymbol{\vec Q}), & i \in \mathcal{M},\ j \in \mathcal{M}, \\
       \hat W_{ij}(\boldsymbol{\vec Q}), & i \in \mathcal{M},\ j \in \mathcal{B}.
    \end{cases}
\end{align}
Note that unlike the purely electronic case, the one-body term can't be directly diagonalized since its coefficients are functions of the nuclear coordinates. Therefore, we partition the Hamiltonian into $\mathcal{N} = N_{\text{mol}} + N_{\text{metal}} + 1$ fragments, $\hat H =\sum_{s=0}^{\mathcal{N}-1} \hat F_s$, such that
\begin{align}
    \hat F_0 =   \sum_{i \in \mathcal{M}} \hat g_{ii}(\boldsymbol{\vec Q}) \hat a_i^\dagger \hat a_i 
    +  \sum_{\substack{i,j \in \mathcal{M}}} \hat V_{ij}(\boldsymbol{\vec Q}) \hat a_i^\dagger \hat a_i \hat a_j^\dagger \hat a_j \,+\,\, \hat U_0 (\boldsymbol{\vec{Q}}),
\end{align}
contains the diagonal elements of the potential with the off-diagonal components partitioned into $\mathcal{N} -2$ fragments
\begin{align} \label{eq:offdiag_general_form}
\sum_{s=1}^{\mathcal{N}-2} \hat F_s  &=  \sum_{s=1}^{\mathcal{N}-2} \sum_{(i,j)\in \mathcal{F}_s} \hat g_{ij}(\boldsymbol{\vec Q}) \left( \hat a_i^\dagger \hat a_j + \hat a_j^\dagger \hat a_i  \right)\nonumber\\ 
&=\sum_{\substack{i \in \mathcal{M} \\ j \in \mathcal{M} \cup \mathcal{B}\\i\neq j}} \hat g_{ij}(\boldsymbol{\vec Q}) \left( \hat a_i^\dagger \hat a_j + \hat a_j^\dagger \hat a_i  \right).
\end{align}
The set of index pairs $\mathcal{F}_s$ in each fragment is given by a matching of the interaction graph. The explicit construction of the matchings and the corresponding Clifford diagonalizing unitaries $\hat D_s = \hat{\mathcal{U}}_s^\dagger \hat F_s \hat{\mathcal{U}}_s$ are described in Appendix~\ref{app:fragmentation}. Lastly, kinetic term is put in its own fragment along with the metal Hamiltonian
\begin{align}
    \hat F_{\mathcal{N}-1} = \sum_{\kappa=0}^{M-1} \frac{\hat P_{\kappa}^2}{2m_{\kappa}} + \sum_{i \in \mathcal{B}} \epsilon_i \hat a_i^\dagger \hat a_i .
\end{align}
We first focus on the fragments corresponding to the potential. Using the Jordan-Wigner mapping we have
\begin{align}\label{eq:diagonal_JW}
\hat D_s 
=
    \begin{cases}\displaystyle
     \sum_{i \in \mathcal{M}} \frac{\hat g_{ii}(\boldsymbol{\vec Q}) }{2}\left(I-\hat \sigma^{(z)}_i\right)
    +  \sum_{\substack{i,j \in \mathcal{M}}} \frac{\hat V_{ij}(\boldsymbol{\vec Q})}{4} \left(I-\hat \sigma^{(z)}_i\right) \left(I-\hat \sigma^{(z)}_j\right) \,+\,\, \hat U_0 (\boldsymbol{\vec{Q}})&, \quad s =0, \\
      \displaystyle \sum_{(i,j) \in \mathcal{F}_s} \frac{\hat g_{ij}(\boldsymbol{\vec Q})}{2} \left( \hat \sigma^{(z)}_{i} - \hat \sigma^{(z)}_{j}  \right)&, \ 1\leq s\leq \mathcal{N}-2.
    \end{cases}
\end{align}
where $\hat \sigma^{(z)}_{i}$ is the Pauli $Z$ operator on qubit $i$.\\

\noindent The task now reduces to implementing the complex exponential of $\hat D_s$ for each fragment. To do this efficiently we expand the nuclear-dependent functions using a set of basis functions $\{\hat f^\lambda(\boldsymbol{\vec Q})\}_\lambda$
\begin{align}
\hat U_{0}(\boldsymbol{\vec Q}) = \,\, \sum_{\lambda} \gamma^{(\lambda)}\, \hat f^\lambda(\boldsymbol{\vec Q}), \quad\quad\hat g_{ij}(\boldsymbol{\vec Q})=  \sum_\lambda \alpha_{ij}^{(\lambda)}\, \hat f^\lambda(\boldsymbol{\vec Q}), \quad\quad\hat V_{ij}(\boldsymbol{\vec Q}) = \, \sum_\lambda \beta_{ij}^{(\lambda)}\, \hat f^\lambda(\boldsymbol{\vec Q}).\nonumber
\end{align}
This allows us to only compute each basis function once per Trotter step and reuse it across fragments. Here we take the basis set to include the monomials from the Taylor expansion of the potentials such as $Q_{\kappa}$, $Q_{\kappa}Q_{\kappa'}$, ..., but to also include analytical functions whose direct computation via QROM \cite{low2024trading} is cheaper than approximating them via a polynomial at relevant grid sizes such as those described in \cref{sec:nuc_potentials}. However, the high-level structure of the algorithm described below is oblivious to the choice of the basis set.\\

\noindent We now re-write \cref{eq:diagonal_JW} in terms of our basis functions
\begin{align}\label{eq:diagonal_expanded}
\hat D_s 
=
    \begin{cases}\displaystyle
     \sum_{\lambda} \hat f^\lambda(\boldsymbol{\vec Q}) \left( \,\sum_{i \in \mathcal{M}} \frac{\alpha_{ii}^{(\lambda)}}{2}\,\left(I-\hat \sigma^{(z)}_i\right)
    +  \sum_{\substack{i,j \in \mathcal{M}}} \frac{\beta_{ij}^{(\lambda)}}{4}, \left(I-\hat \sigma^{(z)}_i\right) \left(I-\hat \sigma^{(z)}_j\right) \,+\,\, \gamma^{(\lambda)}\right)&, \quad s =0, \\
      \displaystyle \sum_\lambda \hat f^\lambda(\boldsymbol{\vec Q})\sum_{(i,j) \in \mathcal{F}_s} \frac {\alpha_{ij}^{(\lambda)}}{2}\,  \left( \hat \sigma^{(z)}_{i} - \hat \sigma^{(z)}_{j}  \right)&, \ 1\leq s\leq \mathcal{N}-2.
    \end{cases}
\end{align}
Notice that when written in this form, the operators can be viewed as a linear combination of basis functions where the coefficients are functions of the state of the electronic register $\ket{\boldsymbol{n}}$. We therefore define
\begin{align}\label{eq:effective_coef}
c_s^{(\lambda)}(\boldsymbol{n})
=
    \begin{cases}\displaystyle
     \,\,\,\sum_{i \in \mathcal{M}} \alpha_{ii}^{(\lambda)}\, n_i
    +  \sum_{\substack{i,j \in \mathcal{M}}} \beta_{ij}^{(\lambda)}\, n_i n_j \,+\,\, h^{(\lambda)}&, \quad s =0, \\
      \displaystyle \sum_{(i,j) \in \mathcal{F}_s} \alpha_{ii}^{(\lambda)}\,(-1)^{n_i} \,(n_i \,\oplus \, n_j)&, \ 1\leq s\leq \mathcal{N}-2.
    \end{cases}
\end{align}
We can then write
\begin{align}\label{eq:diagonal_expanded}
    \hat D_s 
= \sum_\lambda \hat f^\lambda(\boldsymbol{\vec Q}) \cdot c_s^{(\lambda)}(\boldsymbol{n}).
\end{align}
This is the main insight behind our next major optimization, which is the introduction of coefficient accumulation circuits to compute the effective coefficient $c_s^{(\lambda)}(\boldsymbol{n})$ and exponentiate all the terms in a fragment involving $f^\lambda(\boldsymbol{\vec Q})$ simultaneously using a single multiplication to perform the phase gradient operation on the product $f^\lambda(\boldsymbol{\vec Q}) \cdot c_s^{(\lambda)}(\boldsymbol{n})$; this is opposed to using a multiplication for every term in the sum. Here a $b$-qubit phase gradient operation is defined as $\sum_{y=0}^{2^b -1} e^{i 2\pi y/2^b} \ket{y}\bra{y}$, and is performed via modular addition to a register holding the phase gradient state $\ket{R} = \frac{1}{2^{b/2}}\sum e^{i 2\pi y/2^b} \ket{y}$ \cite{gidney2018halving}. Such a phase gradient state only needs to be prepared once at the beginning of the algorithm and is re-used throughout the entire computation.\\

More specifically, given an arbitrary state of the system, $\ket{\Psi} = \sum_{\boldsymbol{n}}\sum_{\boldsymbol{\vec{{x}}}} \psi(\boldsymbol{n},\boldsymbol{\vec{{x}}}) \ket{\boldsymbol{n}} \ket{\boldsymbol{\vec{{x}}}}$, where $\ket{\boldsymbol{n}}$ and $\ket{\boldsymbol{\vec{{x}}}}$ are computational basis states representing the collective state of the electronic and nuclear degrees of freedom correspondingly, we implement $\exp\left(i\hat f^\lambda(\boldsymbol{\vec Q}) \cdot c_s^{(\lambda)}(\boldsymbol{n})\right)$ via the following steps. We begin by computing $f^\lambda(\boldsymbol{\vec x})$ in an ancillary register
\begin{align}
    \ket{\Psi}\ket{0}\ket{0}\ket{R} &= \sum_{\boldsymbol{n}}\sum_{\boldsymbol{\vec{{x}}}} \psi(\boldsymbol{n},\boldsymbol{\vec{{x}}})\, \ket{\boldsymbol{n}} \ket{\boldsymbol{\vec{{x}}}}\ket{0}\ket{0}\ket{R}, \nonumber\\
    &\to \sum_{\boldsymbol{n}}\sum_{\boldsymbol{\vec{{x}}}} \psi(\boldsymbol{n},\boldsymbol{\vec{{x}}})\, \ket{\boldsymbol{n}} \ket{\boldsymbol{\vec{{x}}}}\ket{f^\lambda(\boldsymbol{\vec x})}\ket{0}\ket{R},
\end{align}
by multiplying different nuclear registers if $f^\lambda$ is a monomial, or directly loading it via QROM if $f^\lambda$ is an analytic function. Next, we compute the effective coefficient $c_s^{(\lambda)}(\boldsymbol{n})$ from \cref{eq:effective_coef} in another ancillary register using repeated additions controlled on the parity or product of different qubit pairs in the corresponding fragment $s$
\begin{align}
    \sum_{\boldsymbol{n}}\sum_{\boldsymbol{\vec{{x}}}} \psi(\boldsymbol{n},\boldsymbol{\vec{{x}}})\, \ket{\boldsymbol{n}} \ket{\boldsymbol{\vec{{x}}}}\ket{f^\lambda(\boldsymbol{\vec x})}\ket{0}\ket{R}\to \sum_{\boldsymbol{n}}\sum_{\boldsymbol{\vec{{x}}}} \psi(\boldsymbol{n},\boldsymbol{\vec{{x}}})\, \ket{\boldsymbol{n}} \ket{\boldsymbol{\vec{{x}}}}\ket{f^\lambda(\boldsymbol{\vec x})}\ket{c_s^{(\lambda)}(\boldsymbol{n})}\ket{R}.
\end{align}
Implementation details for computing the effective coefficients $c_s^{(\lambda)}(\boldsymbol{n})$ can be found in \cref{app:coeff}. Lastly, we perform an out-of-place modular multiplication on the registers $\ket{f^\lambda(\boldsymbol{\vec x})}\ket{c_s^{(\lambda)}}$ with the target being the phase gradient register and then uncompute the intermediate results
\begin{align}
    \sum_{\boldsymbol{n}}\sum_{\boldsymbol{\vec{{x}}}} \psi(\boldsymbol{n},\boldsymbol{\vec{{x}}})\, \ket{\boldsymbol{n}} \ket{\boldsymbol{\vec{{x}}}}\ket{f^\lambda(\boldsymbol{\vec x})}\ket{c_s^{(\lambda)}(\boldsymbol{n})}\ket{R} &\to \sum_{\boldsymbol{n}}\sum_{\boldsymbol{\vec{{x}}}} \psi(\boldsymbol{n},\boldsymbol{\vec{{x}}})\, \ket{\boldsymbol{n}} \ket{\boldsymbol{\vec{{x}}}}\ket{f^\lambda(\boldsymbol{\vec x})}\ket{c_s^{(\lambda)}(\boldsymbol{n})}\ket{R \oplus f^\lambda(\boldsymbol{\vec x}) \cdot c_s^{(\lambda)}(\boldsymbol{n})},\nonumber\\
    &= \sum_{\boldsymbol{n}}\sum_{\boldsymbol{\vec{{x}}}} e^{if^\lambda(\boldsymbol{\vec x}) \cdot c_s^{(\lambda)}(\boldsymbol{n})}\, \psi(\boldsymbol{n},\boldsymbol{\vec{{x}}})\, \ket{\boldsymbol{n}} \ket{\boldsymbol{\vec{{x}}}}\ket{f^\lambda(\boldsymbol{\vec x})}\ket{c_s^{(\lambda)}(\boldsymbol{n})}\ket{R}\nonumber\\
    &\to \sum_{\boldsymbol{n}}\sum_{\boldsymbol{\vec{{x}}}} e^{if^\lambda(\boldsymbol{\vec x}) \cdot c_s^{(\lambda)}(\boldsymbol{n})}\, \psi(\boldsymbol{n},\boldsymbol{\vec{{x}}})\, \ket{\boldsymbol{n}} \ket{\boldsymbol{\vec{{x}}}}\ket{0}\ket{0}\ket{R}\nonumber\\
    &= \exp\left(i\hat f^\lambda(\boldsymbol{\vec Q}) \cdot c_s^{(\lambda)}(\boldsymbol{n})\right)\ket{\Psi}\ket{0}\ket{0}\ket{R}.
\end{align}
Finally, putting it all together and incorporating caching techniques, our algorithm proceeds as follows
\begin{mdframed}[
    linewidth=0.5pt,
    linecolor=black!35,
    topline=false,
    bottomline=false,
    rightline=false,
    leftmargin=0pt,
    innerleftmargin=0pt,
    innerrightmargin=0pt,
    innertopmargin=5pt,
    innerbottommargin=5pt
]
\begin{algorithmic}

\For{each $\hat f^\lambda(\boldsymbol{\vec Q})$ in $\{\hat f^\lambda(\boldsymbol{\vec Q})\}_\lambda$}
    \Statex\vspace{-0.2em}

    \If{$\hat f^\lambda(\boldsymbol{\vec Q})$ is a monomial}
        \Statex\vspace{-0.8em}
        \State Compute $\hat f^\lambda(\boldsymbol{\vec Q})$ in an ancillary register using quantum arithmetic
        \Statex\vspace{-0.8em}
    \ElsIf{$\hat f^\lambda(\boldsymbol{\vec Q})$ is an analytical function}
        \Statex\vspace{-0.8em}
        \State Compute $\hat f^\lambda(\boldsymbol{\vec Q})$ in an ancillary register using QROM
    \EndIf

    \Statex\vspace{-0.2em}

    \For{each fragment $0\leq s \leq \mathcal{N}-2$ containing $\hat f^\lambda(\boldsymbol{\vec Q})$}
        \Statex\vspace{-0.8em}
        \State Clifford diagonalize fragment $s$
        \Statex\vspace{-0.8em}
        \State Compute $c_s^{(\lambda)}(\boldsymbol{n})$ in an ancillary register
        \Statex\vspace{-0.8em}
        \State Add the product $\hat f^\lambda(\boldsymbol{\vec Q}) \cdot c_s^{(\lambda)}(\boldsymbol{n})$ to a phase gradient register
        \Statex\vspace{-0.8em}
        \State Uncompute $c_s^{(\lambda)}(\boldsymbol{n})$
        \Statex\vspace{-0.8em}
        \State Clifford undiagonalize fragment $s$
    \EndFor

    \Statex\vspace{-0.6em}

    \State Uncompute $\hat f^\lambda(\boldsymbol{\vec Q})$
\EndFor

\end{algorithmic}
\end{mdframed}
Lastly, the fragment $\hat F_{\mathcal{N}-1} = \sum_{\kappa=0}^{M-1} \frac{\hat P_{\kappa}^2}{2m_{\kappa}} + \sum_{i \in \mathcal{B}} \epsilon_i \,\hat a_i^\dagger \hat a_i$ is exponentiated using $R_Z(\epsilon_i)$ rotations on the qubits of the metal register and the implementation of $\exp{i\sum_{\kappa=0}^{M-1} \frac{\hat P_{\kappa}^2}{2m_{\kappa}}}$ is done via squaring circuits and phase gradient operations after diagonalization via QFT by \cref{eq:qft_diagonalization}.

\section{Inputs and Outputs} \label{sec:i/o}

\subsection{Initial states}
Meaningful simulation results depend critically on the preparation of initial states that accurately reflect the underlying physical conditions. Here we assume an initial condition in which the molecule and metal are sufficiently separated/weakly coupled such that the total state \emph{initially} can be well represented as
\begin{align} \label{eq:initial_state}
\hat \rho(0) = \ket{\Psi(0)}\bra{\Psi(0)}_{\rm{mol}}  \otimes \hat \rho_{\rm{metal}} (0).
\end{align}
For many practical instances the initial state of the molecule can written in the product form
\begin{align}\label{eq:initial_state_mol}
    \ket{\Psi(0)} = \ket{\Phi}_{el} \bigotimes_{\kappa} \ket{\chi}_\kappa,
\end{align}
where $\ket{\Phi}$ is electronic ground state of the molecule, and $\ket{\chi}_\kappa$ would be the harmonic oscillator ground state if $\kappa$ is a non-reactive mode, the Morse oscillator ground state if it's a reactive mode, and a Gaussian wavepacket with an initial velocity towards the surface if it's a translational mode.\\

\noindent We assume the metal to initially be in a thermal equilibrium given by grand canonical thermal state
\begin{align} \label{eq:gibbs_state}
\hat \rho_{\rm{metal}}(0) = \frac{1}{\mathcal{Z}} \exp\left(-\sum_{i \in \mathcal{B}} \frac{(\epsilon_{i} -\mu)}{k_BT}\, \hat a_{i}^\dagger \hat a_{i}\right),
\end{align}
where $\mu$ is the chemical potential, $k_B$ is the Boltzmann constant, and $T$ is the temperature. Due to the separability of the metal electronic DOFs, preparing $\hat \rho_{\rm{metal}} (0)$ amounts to probabilistically initializing metal orbital qubits as $\{\ket{0}_k$, $\ket{1}_k$\} according to independent Bernoulli distributions
\begin{align}\label{eq:initial_state_metal}
\hat \rho_{\rm{metal}}(0) = \bigotimes_{k \in \mathcal{B}} \Big( (1 - f_k) \ketbra{0}{0}_k + f_k \ketbra{1}{1}_k \Big), \quad  f_k = \frac{1}{1 + e^{(\epsilon_k - \mu) / k_B T}}. 
\end{align}

\subsection{Observables}

\subsubsection{Electronic observables}
The time-dependent occupations of a molecules electronic orbitals are directly related to charge transfer dynamics at the interface, providing a measure of how rapidly electrons are exchanged between the molecule and metal substrate. Such dynamical information is central to understanding interfacial electron transfer rates, and holds promise in enabling the discovery of functional materials with enhanced efficiency \cite{liu2017two}. The electronic state populations of the molecule are given by 
\begin{align} \label{eq:electronic_populations}
P_i(t) = \text{Tr}\left( \hat \rho(t) \hat a_i^\dagger \hat a_i \right) = \text{Tr}\left(\hat \rho(t) \ket{1}\bra{1}_i\right),
\end{align}
for $i \in \mathcal{M}$, and number of electrons in the molecule can be tracked as $P_{\rm{el}}(t) = \sum_{i \in \mathcal{M}} P_i(t)$, amounting to measuring the molecular electronic register in the computational basis, where simultaneous estimation of all $N_{\rm{mol}}$ occupation numbers is achieved using $\mathcal{O}(\epsilon^{-2})$ measurements for an $\epsilon$ accurate estimation. Alternatively, one could use quantum amplitude estimation to directly estimate $P_{i}(t)$ with $\mathcal{O}(\epsilon^{-1})$ complexity at the cost of longer circuit depth \cite{brassard2000quantum}.

\subsubsection{Nuclear observables}
An important observable on the molecular nuclear DOFs is their time-dependent positions. The position of nuclear coordinate $Q_{\kappa}$ at time $t$ is
\begin{align} \label{eq:avg_pos}
Q_{\kappa}(t) = \text{Tr}\left( \hat \rho(t) \hat Q_{\kappa} \right) = \sum_{p=0}^{K - 1} x_p \text{Tr} \left( \hat \rho(t) \ket{x_p} \bra{x_p}  \right),
\end{align}
and can be obtained by measuring the $\kappa^{\rm{th}}$ nuclear DOF register in the computational basis, providing distribution $\{ \text{Tr} \left( \hat \rho(t) \ket{x_p} \bra{x_p}  \right) \}_{p=0}^{K-1}$. \cref{eq:avg_pos} provides the average, but other observables of interest could be constructed from this distribution. For instance, one could bipartition a reactive or translational coordinate's grid into a \textit{reactant} $\Omega_R$ and \textit{product} $\Omega_P$ region and recover time-resolved populations within these complement subspaces to track processes such as molecular bond dissociation or surface adsorption. One could also in principle use projector $\hat \Omega_{P} = \sum_{x_p \in \Omega_{P}} \ket{x_p}\bra{x_p}$ in a flux correlation function approach to extract a thermal rate of reaction \cite{miller1983quantum, wang2006calculation}. \\

Another critical observable is the time-resolved vibrational \textit{energy} populations, which provide a probe to non-adiabatic energy flow at the molecule-metal surface. Such information reveals how specific nuclear DOFs exchange energy with electron-hole pairs, which can elucidate dissipation pathways and energy transfer dynamics, such as in molecular beam scattering and vibrationally activated reactions at catalytic surfaces \cite{jiang2012enhancing, rittmeyer2018energy, crim2008chemical, shaon2025tuning}. The vibrational energy distribution of nuclear coordinate $\kappa$ is given by
\begin{align} \label{eq:vib_energy}
P_{\nu}^{(\kappa)} (t) = \text{Tr} \left( \hat \rho(t) | \phi_{\nu}^{(\kappa)} \rangle \langle \phi_{\nu}^{(\kappa)} |  \right),
\end{align}
where $|\phi_{\nu}^{(\kappa)} \rangle$ is the $\nu^{\rm{th}}$ vibrational level. A suitable choice of basis must be made, such as the unimodal vibrational eigenstates in the reference nuclear potential $\hat U_0(\boldsymbol{\vec{Q}})$. For modes that are predominantly harmonic, \cref{eq:vib_energy} can be estimated via computational basis measurements after a basis transformation from position basis to harmonic oscillator eigenstates using a quantum Hermite transform \cite{jain2025efficient}. For anharmonic modes, the necessary basis transformation has matrix elements $U_{pp'} = \langle x_p | \phi_{p'} \rangle = \phi_{p'}(x_p)$ and could in general be carried out using unitary synthesis techniques \cite{wierichs2025unitary}.

\section{Applications} \label{sec:applications}

\newcommand{\softsection}[1]{\vspace{0.7em}\noindent\textbf{#1}\quad}

Non-adiabatic dynamics of molecules interacting with metal surfaces govern a remarkably wide range of processes. Because these processes frequently lie in regimes where classical methods are limited by either computational cost or uncontrolled approximations, they offer compelling opportunities for quantum simulation. Within this section, we highlight representative areas in which the quantum algorithm presented in this work could be applied to dynamical simulations of technological importance. We report quantum resource estimation for $1000$ first-order Trotter steps under model systems reflective of these applications in \cref{tab:resource_estimation}. Details on the representative models are included in \cref{app:models}, along with a discussion on initial states and observables associated to each simulation task, illustrating how the algorithm could be applied in practice. \\

\softsection{Molecular scattering and adsorption} \newline

The non-adiabatic dynamics of molecules scattering and adsorbing at metallic surfaces govern interfacial processes across many settings of practical significance. In heterogeneous catalysis, inelastic scattering and adsorption are elementary steps determining how incident molecules exchange energy with the surface and access reactive configurations. Examples of such dynamics include the activation of H$_2$ and O$_2$ on metal surfaces, involved in key catalytic processes such as ammonia synthesis \cite{ertl2008reactions}, hydrocarbon hydrogenation \cite{sabatier1922catalysis}, and CO oxidation \cite{min2007heterogeneous}. Non-adiabatic effects in these processes have been directly observed \cite{wodtke2004electronically, carbogno2013o2}, and recent experiments \cite{borodin2022quantum} demonstrate that nuclear quantum effects can be significant even at $250–1000$ K, i.e., nominal temperatures of industrial catalysis operating environments. Under such conditions, classical and harmonic transition-state theories overestimated reaction rates by up to $10^3$ fold \cite{borodin2022quantum}, suggesting that there are industrially relevant regimes for which understanding and predicting catalytic reactions would benefit substantially from quantum dynamical simulations. \\

Another key setting where non-adiabatic dynamics arise is during adsorbate motion at electrochemical interfaces, such as during electrochemical proton transfer \cite{pecina1998model, navrotskaya2008model} and coupled ion-electron transfer \cite{bazant2023unified}. Electron transfer rates can depend strongly on adsorbate nuclear motion and solvent reorganization, requiring a description which takes the coupled electronic-nuclear dynamics into account \cite{ouyang2015surface, arguelles2024time}. For resource estimation, we consider a GAN Hamiltonian representative of a complex electron-transfer induced chemical process occurring as a result of reactant motion towards an electrode surface. Such electrochemical interfacial dynamics are central to industrially important settings ranging from battery electrodes to metal interfaces susceptible to corrosion \cite{katz1999non, behler2008nonadiabatic, shimakura2025quantum}, where chemical processes involving non-adiabatic charge-transfer can have a decisive role in performance and stability. \\ 

\renewcommand{\arraystretch}{1.25} 
\begin{table}[t!]
    \begin{ruledtabular}
    \begin{tabular}{lcccccc}
        \textbf{Simulation task} & 
        $N_{\rm{mol}}$ & 
        $N_{\rm{metal}}$ & 
        $M$ & 
        $(M', M{''}, M_{\rm{z}})$ &
        \textbf{Qubits} &
        \textbf{Toffoli gates} \\
        \colrule
        \multirow{3}{*}{\shortstack{Molecular scattering/adsorption}} & 
        $4$ & $50$ & $14$ & ($10$, $3$, $1$) & $196$ & $2.30 \times 10^{7}$ \\
        & $4$ & $100$ & $14$ & ($10$, $3$, $1$) & $246$ & $3.05 \times 10^{7}$ \\
        & $4$ & $100$ & $34$ & ($30$, $3$, $1$) & $346$ & $4.55 \times 10^{7}$ \\
        \colrule
        \multirow{3}{*}{\shortstack{Photoinduced charge transfer}} & 
        $4$ & $100$ & $20$ & ($20$, $0$, $0$) & $267$ & $2.02 \times 10^7$ \\
        & $8$ & $100$ & $20$ & ($20$, $0$, $0$) & $271$ & $7.85 \times 10^7$ \\
        & $8$ & $100$ & $40$ & ($40$, $0$, $0$) & $371$ & $1.29 \times 10^8$ \\
        \colrule
        \multirow{3}{*}{\shortstack{Molecular junction transport}} & 
        $2$ & $100$ & $8$ & ($5$, $2$, $1$) & $212$ & $2.79 \times 10^7$ \\
        & $2$ & $100$ & $13$ & ($10$, $2$, $1$) & $237$ & $3.10 \times 10^7$ \\
        & $2$ & $200$ & $13$& ($10$, $2$, $1$) & $337$ & $5.02 \times 10^7$ \\
        \colrule
    \end{tabular}
    \end{ruledtabular}
    \caption{\label{tab:resource_estimation}%
    Estimated cost of our quantum algorithm for representative molecule-metal systems at various levels of complexity. The reported numbers correspond to taking $1000$ Trotter steps, which are likely sufficient for many practical applications. A full analysis on the number of required Trotter steps is outside the scope of this paper. We use $M'$, $M''$, and $M_z$ to denote the number \textit{non-reactive}, \textit{reactive}, and \textit{translational} modes discretized into $32$, $128$, and $256$ gridpoints correspondingly. It is worth noting that our resource estimates do not take into account the sparsity present in real-world systems which can further reduce the costs presented here. The code used to calculate the presented estimates uses the Pennylane \cite{bergholm2018pennylane} estimator module~\cite{soni2026howto}. The code is available in \cite{GAN-dynamics}.} 
\end{table}
\renewcommand{\arraystretch}{1.0}

\newpage 
\softsection{Photoinduced heterogeneous charge transfer} \newline 

Another area where non-adiabatic interfacial dynamics play a central role is in photoinduced heterogeneous charge transfer at molecule–semiconductor interfaces, in which light absorption drives electron transfer across the interface in interfacial photovoltaic systems \cite{swierk2016ultrafast,brennaman2016finding,house2015artificial}. Dye-sensitized solar cells represent a concrete technologically important example \cite{o1991low,hagfeldt2010dye}, where photoexcited visible light absorbers inject electrons into the conduction band of a semiconducting material in the ultrafast regime.  The typical timescale of photoinduced charge injection for dye--semiconductor systems is on the order of $10$ to $100$ femtoseconds, where it has been indicated experimentally and theoretically that coupling to vibrational modes of the dye plays a significant role in the charge transfer dynamics \cite{thoss2007quantum, zimmermann2001experimental, willig2000ultrafast, li2015quantum}.\\ 

The rate of photoinduced charge transfer is a decisive factor in the performance of interfacial optoelectronic devices, as it determines how efficiently photoexcited carriers are converted into separated charges, driving the production of photocurrent. Consequently, the ability to extract accurate non-adiabatic charge-injection rates through quantum dynamical simulations would provide a powerful tool for rational material design for dye-sensitized solar cells \cite{liu2017two}. However, equally important is accounting for loss channels that occur on comparable timescales, including internal conversion and back electron transfer from the semiconductor conduction band to the oxidized dye \cite{anderson2004ultrafast,anderson2005ultrafast}. We assess the resource requirements of the quantum algorithm using a GAN Hamiltonian parameterization which in principle captures both initial charge injection and the dominant loss mechanisms to provide a benchmark for a challenging case of photoinduced interfacial electron transfer processes. \\ 

\softsection{Charge transport across molecular junctions} \newline 

A further application of non-adiabatic quantum dynamical simulations at metal interfaces arises in molecular junctions, in which a single molecule or small molecular assembly connects two metallic leads and serves as the minimal functional unit of molecular electronics. These systems offer the prospect of ultra–low-power electronics, chemically tunable transport characteristics, and novel functionalities \cite{gupta2023nanoscale, aradhya2013single}. However, their technological viability is limited by the poor stability of molecules under sustained voltage bias, where electrical currents drive vibrational excitation and structural degradation. Understanding and mitigating these failure mechanisms is therefore a central challenge in designing reliable molecular electronic components. Furthermore, their simulation provides a valuable probe into current-induced chemistry more broadly \cite{jorn2010implications, dzhioev2013out}. \\ 

The underlying out-of-equilibrium dynamical processes involve the coupling between molecular nuclear degrees of freedom, the molecule’s internal electronic states, and the electronic continua of the two metallic leads. This interaction makes simulations of molecular junction dynamics a strongly non-adiabatic quantum transport problem. We perform resource estimation for a GAN Hamiltonian describing a complicated instance of current-induced bond rupture \cite{erpenbeck2018current, ke2021unraveling, li2016mechanism}, in which electronic currents can lead to vibrational heating \cite{koch2006current, hartle2011resonant} and transient population of antibonding electronic states \cite{erpenbeck2020current}, driving the dissociation of chemical bonds. 

\section{Discussion and Outlook} \label{sec:discussion}

In this work, we have presented the first quantum algorithm to address the non-adiabatic dynamics of molecules at metal surfaces via time evolution under a generalized Anderson-Newns (GAN) Hamiltonian. The model incorporates an arbitrary number of molecular and metal electronic states and molecular nuclear DOFs. Further, in its full generality, the model captures anharmonic vibronic interactions, geometry-dependent molecule-metal coupling, and electronic correlation effects. Since the algorithm performs the unitary dynamics in the full molecule–metal Hilbert space, the evolution is completely non-perturbative, avoiding assumptions of weak coupling, Markovian fermionic bath, or classical nuclear motion, thereby providing access to high fidelity time-resolved observables in regimes that are inaccessible to classical computing methods. Moreover, the quantum algorithm makes use of a number of techniques to reduce the implementation cost of time evolution using quantum arithmetic and phase gradients. The use of coherent coefficient accumulation allows for only a single phase gradient operation per basis function within a fragment, resulting in a drastic reduction when compared to performing the phase gradient for each term that the basis function appears in. Furthermore, utilization of caching techniques allows each basis function to only be computed once per Trotter step,  substantially reducing the amount of redundant arithmetic. Lastly, by identifying basis functions which are poorly represented by low degree polynomials, we see significant cost savings in practice by opting to QROM \cite{low2024trading} the function values rather than computing them using arithmetic. The combination of these algorithmic components yields a markedly low-cost implementation of each Trotter step. \\ 

While the present framework captures a remarkably broad range of interfacial physics, it also admits several natural extensions that could further enhance its predictive power and domain of applications. In principle, phononic DOFs of the metal substrate or a solvent environment can already be incorporated directly as additional explicit nuclear DOFs within the GAN considered in this work, although in practice they may possibly be treated more efficiently through system–bath formulations. Another possible extension is to incorporate rotational DOFs of the molecule, which would enable the description of orientation-dependent coupling to the metal, anisotropic scattering dynamics, and rovibrational energy transfer. A critical future aspect is that of accurately parameterizing the GAN Hamiltonian. The construction of the model ultimately relies on electronic structure calculations to parameterize energies, couplings, and hybridization with the metal, a problem outside the scope of the present work. However, classical electronic structure methods for interfacial systems are advancing rapidly \cite{meng2022pragmatic, wu2024constrained, chen2025extremely, qiu2025fast}, and we anticipate that these developments will be instrumental in reliable parameterization of the model for future applications. \\ 

This work constitutes a concrete step toward a broader objective: establishing simulations of real-time quantum dynamics as the natural and industrially relevant application domain for quantum computers, in contrast to the predominantly explored paradigm of static energy calculations. Dynamical problems offer a regime in which meaningful quantum advantage can be achieved at smaller system sizes, owing to the substantially larger performance gap between quantum and classical approaches. This separation directly translates into reduced resource requirements for practically relevant applications. We hope this work motivates increased focus within the community on quantum dynamics as a central application of quantum computing.

\section{Acknowledgments}
The authors thank Stepan Fomichev, Alain Delgado, Shreyas Malpathak, and Ignacio Loaiza for stimulating discussions, and Tarik El-Khateeb for producing the \cref{fig:hero} graphic. We acknowledge the Applied Quantum Computing Challenge program of the National Research Council of Canada for financial support (grant number AQC-103-2).

\bibliography{main}

\begin{appendix}

\section{Fragmentation Scheme} \label{app:fragmentation}

To identify fragments of mutually commuting terms, it is convenient to work in the basis of Majorana fermion operators defined by $\hat \gamma_{2i} = \hat a_i + \hat a_i^\dagger$ and $\hat \gamma_{2i+1} = - i ( \hat a_i - \hat a_i^\dagger )$. In this basis, the off-diagonal hopping terms in the Hamiltonian take the form
\begin{align} \label{eq:h_mol_metal_maj}
\hat a_i^\dagger \hat a_j + \hat a_j^\dagger \hat a_i = \frac{i}{2} \left( \hat \gamma_{2i} \hat \gamma_{2j+1} + \hat \gamma_{2j} \hat \gamma_{2i+1}  \right).
\end{align}
Two terms $\hat \gamma_\alpha \hat \gamma_\beta$ and $\hat \gamma_\mu \hat \gamma_\nu$ commute if and only if they share either zero or two indices. Consequently, a set of hopping terms mutually commute if the corresponding indices $(i, j)$ form a \textit{matching} on the interaction graph, that is, a set of edges where no two edges share a common vertex. For a fragment defined by such a matching $\mathcal{F}_s$, the operator is written as
\begin{align}
\hat F_s = \sum_{(i,j) \in \mathcal{F}_s} \frac{i \hat g_{ij}(\boldsymbol{\vec Q})}{2} \left( \hat \gamma_{2i} \hat \gamma_{2j+1} + \hat \gamma_{2j} \hat \gamma_{2i+1}  \right).
\end{align}
This operator consists of mutually commuting degree $2$ Majorana products, and hence can be diagonalized by fermionic matchgate unitaries which are also Clifford \cite{wan2023matchgate}, taking the form 
\begin{align} \label{eq:clifford_so2n_unitary}
\hat{\mathcal{U}}_s = \prod_{(i,j) \in \mathcal{F}_s} e^{-\frac{\pi}{4} \hat \gamma_{2j+1} \hat \gamma_{2i+1}},
\end{align}
which in the Jordan-Wigner representation yields the diagonal form presented in \cref{eq:diagonal_JW}, $\hat D_s = \hat{\mathcal{U}}_s^\dagger \hat F_s \hat{\mathcal{U}}_s = \sum_{(i,j) \in \mathcal{F}_s} \hat g_{ij}(\boldsymbol{\vec Q}) ( \hat \sigma^{(z)}_{i} - \hat \sigma^{(z)}_{j} ) / 2$.\\

\noindent To cover all terms in the Hamiltonian, we decompose the edge set of the interaction graph into disjoint matchings $\bigcup_{s} \mathcal{F}_s = \mathcal{E}$. For the molecular interactions ($\hat H_{\text{mol}}$), the graph is fully connected with $N_{\text{mol}}$ vertices. If $N_{\text{mol}}$ is even, we utilize the canonical 1-factorization of $K_{N_{\text{mol}}}$, requiring $N_{\text{mol}} - 1$ fragments. If $N_{\text{mol}}$ is odd, we extend the graph to include a dummy node and apply the even case factorization, resulting in $N_{\text{mol}}$ fragments. For the even case, the explicit matchings are given by
\begin{align}
\mathcal{F}_s^{(\text{mol})} = \left\{ \left( m_{[(s+k) \text{ mod } (N_{\rm{mol}}-1)]}, m_{[(s-k) \text{ mod } (N_{\rm{mol}}-1)]} \right) : k = 1, 2,\dots, \frac{N_{\text{mol}}}{2} - 1 \right\} \cup \{(m_s, m_{N_{\text{mol}}-1})\}.
\end{align}
For the metal-molecule interactions ($\hat H_{\text{int}}$), the graph is complete bipartite with partitions of size $N_{\text{mol}}$ and $N_{\text{metal}}$. Assuming $N_{\text{mol}} \le N_{\text{metal}}$, the edges can be decomposed into $N_{\text{metal}}$ matchings defined by the cyclic shifts
\begin{align}
\mathcal{F}_s^{(\text{int})} = \{(m_i, b_{[(i+s) \text{ mod } N_{\text{metal}}]}) : i = 0, 1,\dots, N_{\text{mol}}-1 \}.
\end{align}

\section{Coefficient Accumulation Circuits} \label{app:coeff}

We now describe the circuit implementation for coherently computing the effective coefficients $c_s^{(\lambda)}(\boldsymbol{n})$ defined in \cref{eq:effective_coef}. The goal is to accumulate the target value into a $c$-bit register $\ket{C}$ initialized to zero, $\ket{\boldsymbol{n}}\ket{0}_C \mapsto \ket{\boldsymbol{n}}\ket{c_s^{(\lambda)}(\boldsymbol{n})}_C$. For the diagonal fragment ($s=0$), the coefficient is a quadratic function of the occupation numbers
\begin{align}
c_0^{(\lambda)}(\boldsymbol{n}) = \sum_{i \in \mathcal{M}} \alpha_{ii}^{(\lambda)} n_i + \sum_{i,j \in \mathcal{M}} \beta_{ij}^{(\lambda)} n_i n_j + h^{(\lambda)}.
\end{align}
Since $n_i \in \{0, 1\}$, this is computed by strictly classical arithmetic controlled by the quantum state. For each non-zero term $\beta_{ij}^{(\lambda)}$, we load the constant $\beta_{ij}^{(\lambda)}$ into an ancilla register conditioned on $n_i \cdot n_j = 1$ (using a Toffoli), add the ancilla to the accumulator $\ket{C}$, and uncompute the ancilla. This process requires $\mathcal{O}(N_{\text{mol}}^2)$ controlled additions.\\

\noindent For off-diagonal fragments ($s \neq 0$), the effective coefficient depends on the parity of the occupations involved in the matching $\mathcal{F}_s$:
\begin{align} \label{eq:app_coeff_target_s}
c_s^{(\lambda)}(\boldsymbol{n}) = \sum_{(i,j) \in \mathcal{F}_s} \alpha_{ij}^{(\lambda)}\,(-1)^{n_i} \,(n_i \oplus n_j).
\end{align}
The term for pair $(i,j)$ is non-zero only if $n_i \neq n_j$, with the sign determined by $n_i$. We implement this by first computing the parity $p_{ij} = n_i \oplus n_j$ into a single ancilla qubit using CNOTs. Conditioned on $p_{ij}=1$, we load the constant coefficient $\alpha_{ij}^{(\lambda)}$ into a $c$-bit ancilla register. If $n_i=1$, we negate the value in the ancilla register to apply the $(-1)^{n_i}$ factor. We then add the signed value to the accumulator $\ket{C}$ and uncompute the ancillas in reverse order.

\section{Model specifications} \label{app:models}
Here we describe in detail the model specifications utilized for the resource estimations reported in \cref{tab:resource_estimation}. For all systems, unless otherwise specified, the following parameterizations are chosen.
In $\hat H_{\rm{mol}}$, the nuclear coordinate-dependent couplings and potentials are given by the forms
\begin{align} \label{eq:RE_U_ij}
\hat U_{ij}(\boldsymbol{\vec Q})
&=
\begin{cases}
\displaystyle \lambda^{(i,j)} + \sum_{\kappa = 0}^{M-1} \,a_{\kappa}^{(i,j)} \hat Q_{\kappa} +  b_{\kappa \kappa}^{(i,j)} \hat Q_{\kappa}^2  & i\neq j\,\,\\[15pt]
\displaystyle \lambda^{(i)} + \sum_{\kappa = M'}^{M-1} f_{\kappa}^{(i)}(\hat Q_{\kappa}) +  \sum_{\kappa = 0}^{M'-1} \,a_{\kappa}^{(i)} \hat Q_{\kappa} + b_{\kappa \kappa}^{(i)} \hat Q_{\kappa}^2 \qquad \qquad \qquad  & i=j\,\,,
\end{cases}
\end{align}
and $\hat U_0(\boldsymbol{\vec{Q}})$ takes the same form as the $i = j$ case. Similarly, the vibronic contribution to the electron-electron interaction is assumed to take the form
\begin{equation} \label{eq:RE_V_ij}
    \hat V_{ij}(\boldsymbol{\vec Q}) = h^{(ij)} + \sum_{\kappa = 0}^{M-1} \,c_{\kappa}^{(ij)} \hat Q_{\kappa} + d_{\kappa}^{(ij)} \hat Q_{\kappa}^2.
\end{equation}
In $\hat H_{\rm{int}}$, all $\hat W_{ij} (\boldsymbol{\vec{Q}})$ depend only on the molecule-metal distance coordinate, and they possess the same functional dependence on the distance coordinate, $\hat W_{ij}(\boldsymbol{\vec{Q}}) = w_{ij} \hat f_{\rm{int}}(Q_{M-1})$, where $\hat f_{\rm{int}}(Q_{M-1})$ is an analytical expression such as in \cref{eq:W}. This assumption is commonly justified as the molecule–metal couplings typically decay with distance on similar length scales, and their relative strengths are sufficiently encoded in the coefficients $w_{ij}$. Importantly, this form does not invoke the wide-band approximation, and the resulting energy- and coordinate-dependent hybridization can generate memory effects requiring system-bath descriptions of the dynamics to go beyond the Markovian approximation, which further complicates classical simulations. Since in all resource estimates reported, the analytical nuclear functions are directly included in the basis set and implemented via QROM, their cost is independent of the specific functional form used, and instead depends on the size of the underlying real-space grid. Hence, we do not need to consider explicit forms for the analytical expressions for the purposes of obtaining their cost of implementation. Furthermore, we assume that all coefficients entering potentials \cref{eq:RE_U_ij,eq:RE_V_ij} are non-zero, and hence the Trotter resource estimations reported in \cref{sec:applications} should be considered as upper bounds to models of the same size specifications that have been parameterized from ab initio data, which could potentially benefit substantially from sparsity. We emphasize that the models presented here are not intended to describe specific molecule-metal systems, but have been devised strictly for the purpose of performing resource estimation of the quantum algorithm for problem instances representative of the three applications discussed. However, we provide sketches of possible initial states and observables of interest to illustrate how the algorithm applied to such models could be used in practice to extract dynamical information relevant to these application domains.

\subsection{Molecular scattering/adsorption}
To assess the cost of the quantum algorithm applied to a complex electrochemical reaction occurring during adsorbate motion towards an electrode surface, we consider a spinful GAN Hamiltonian resolved in a basis of $N_{\rm{mol}} = 4$ spin orbitals, which could represent the two frontier unoccupied spatial orbitals local to the molecule which are in energetic proximity to the electrode Fermi level. The potential $\hat U_0 (\boldsymbol{\vec{Q}})$ represents the ground state potential energy surface of the neutral species. Explicitly, we have
\begin{align} \label{eq:h_mol_spinful}
\hat H_{\rm{mol}} = \sum_{\kappa = 0}^M \frac{\hat P_\kappa^2}{2 m_{\kappa}} + \hat U_0 (\boldsymbol{\vec{Q}}) + \sum_{\substack{i,j=0 \\ \sigma}}^1 \hat U_{ij} (\boldsymbol{\vec{Q}}) \hat a_{i \sigma}^\dagger \hat a_{j \sigma} + \sum_{\substack{i,j = 0 \\ \sigma, \sigma'}}^1 \hat V_{ij} (\boldsymbol{\vec{Q}}) \hat a^\dagger_{i \sigma} \hat a_{i \sigma} \hat a^\dagger_{j \sigma'} \hat a_{j \sigma'},
\end{align}
where $\sigma \in \{ \alpha, \beta \}$ are the orbital spin-z projections. Note that in the spinful representation, $\hat H_{\rm{metal}}$ and $\hat H_{\rm{int}}$ are defined as 
\begin{align} \label{eq:h_metal_spinful}
\hat H_{\rm{metal}} = \sum_{\substack{i \in \mathcal{B} \\ \sigma}} \epsilon_i \hat a_{i\sigma}^\dagger \hat a_{i \sigma},
\end{align}
and 
\begin{align} \label{eq:h_int_spinful}
\hat H_{\rm{int}} = \sum_{\substack{i \in \{ 0,1  \} \\ j \in \mathcal{B} \\ \sigma}} \hat W_{ij}(\boldsymbol{\vec{Q}}) \left( \hat a^\dagger_{i \sigma} \hat a_{j \sigma} + \hat a^\dagger_{j \sigma} \hat a_{i \sigma} \right),
\end{align}
as they are also spin conserving operators in the absence of spin-orbit coupling.
For the nuclear DOFs, the translational coordinate describes the distance between the molecular system and the electrode surface, and we consider $3$ reactive internal modes. For instance, the molecular system could be describing a three-fold coordinated solvation complex, with reaction coordinates representing the solvent-solute bond distances. Further, we consider up to $30$ non-reactive vibrational modes, which could represent selected explicit vibrational coordinates of the solvation complex, and/or a set of bosonic modes emulating a desired bath spectral density function to capture implicit solvent effects. 

\subsubsection*{Initial states}
To simulate the dynamics occurring during an electrochemical adsorption process, the state \cref{eq:initial_state} describing an initially neutral adsorbate with an initial velocity towards the surface would utilize
\begin{align}
\ket{\Psi(0)}_{\rm{mol}} = \bigotimes_{i=0}^{3}\ket{0}_i \bigotimes_{\kappa=0}^{M-1} \ket{\chi}_{\kappa},
\end{align}
where, for non-translational nuclear DOFs ($Q_{\kappa}$ for $\kappa \in \{0, 1, \hdots, M-2\}$), state $\ket{\chi}_{\kappa}$ could be initialized to their Morse or harmonic oscillator ground states, depending on if they are reactive or non-reactive, and the translational DOF $Q_{M-1}$ can be initialized as a Gaussian wavepacket with kinetic energy $p^2/2m_{\kappa}$, width $\sigma$, and centered at $q_0$,
\begin{align}
\chi (Q_{M-1}) = \mathcal{N} \exp \left( - \frac{\left(Q_{M-1} - q_0  \right)^2}{4 \sigma^2} + i p \left( Q_{M-1} - q_0  \right)  \right),
\end{align}
where $\mathcal{N}$ is a normalization factor. The initial state of the metal at temperature $T$ is given by \cref{eq:initial_state_metal}. For electrochemical systems in the $T \gg 0$ regime, it is likely of importance to thermalize the molecular DOFs as well, though initializing both the molecular and metal DOFs as pure states is justified in the low temperature regime where quantum effects are expectedly most pronounced. 

\subsubsection*{Observables}

Various observables of interest could be extracted along the time evolution of the adsorption dynamics. For instance, in a reductive adsorption process, the electron transfer rate from the metal substrate to the adsorbate species is of crucial importance, which can be extracted by tracking the electronic populations of the molecular spin orbitals over time, 
\begin{align} \label{eq:electronic_population_spinful}
P_{\rm{el}}(t) = \sum_{i, \sigma} P_{i \sigma}(t),
\end{align}
where $P_{i \sigma}(t)$ is given in \cref{eq:electronic_populations}. Simultaneously, observables on the nuclear DOFs can be sampled, including the average position of the translational coordinate $Q_{M-1}$ using \cref{eq:avg_pos}. Partitioning the $Q_{M-1}$ grid into \textit{adsorbed} and \textit{unadsorbed} subspaces, a rate of adsorption can also be obtained by evaluating the expectation value of projector
\begin{align} \label{eq:ads_projector}
\hat \Omega_{\rm{ads}} = \sum_{x_p \in \Omega_{\rm{ads}}} \ketbra{x_p}{x_p}_{M-1}
\end{align}
over time, where $\Omega_{\rm{ads}}$ is the set of gridpoints defining the adsorbed subspace, and its expectation value is straightforwardly obtained via computational basis measurements on the $(M-1)^{\rm{th}}$ nuclear DOF register in the same manner as \cref{eq:avg_pos}. Similarly, the position distributions of the three internal reactive coordinates can also be simultaneously sampled to dynamically track dissociation/desolvation as the system approaches the metal surface. Collectively, time-resolved tracking of these observables would deliver direct insight into the interplay between electron transfer, desolvation/dissociation, and surface adsorption, which could provide valuable mechanistic understanding relevant to the rational design of corrosion inhibitors, battery electrolytes, and electrode materials.

\subsection{Photoinduced heterogeneous charge transfer}
To model the heterogeneous charge transfer at the dye-semiconductor interface, we again consider a GAN resolved in the spin orbital basis, with the general form the same as described in \cref{eq:h_mol_spinful,eq:h_metal_spinful,eq:h_int_spinful}. Here, the molecular electronic DOFs are described by the spin orbitals corresponding to the highest occupied molecular orbitals (HOMO) and lowest unoccupied molecular orbitals (LUMO) localized to the dye. For instance, the $N_{\rm{mol}} = 4$ case could utilize the HOMO and LUMO space, and the $N_{\rm{mol}}=8$ case could use the HOMO$-1$, HOMO, LUMO, LUMO$+1$ space. In describing the metal, we assume inclusion of both the valence band and conduction band orbitals. We note that by inclusion of multiple occupied spin orbitals of the dye and the populated valence band of the metal, the assumed form represents a many-electron problem. This is in contrast to previous theoretical studies of the non-adiabatic quantum dynamics of photoinduced heterogeneous charge transfer which consider strictly one-electron donor and acceptor states, for which standard vibronic Hamiltonians can be employed \cite{liu2017isomorphism}. The conceptual model described here would, in principle, account for electron correlation effects between electrons localized at the dye, the role of multiple donor states, and the competition between charge injection and hole transfer from the valence band to the dye, which collectively would provide a more complete description of the charge transfer dynamics. Furthermore, the model concept used here would account for internal conversion within the dye, which could compete with charge injection. \\

To characterize the nuclear coordinates, we consider all DOFs to be non-reactive, representing the vibrational modes of the dye. Additionally, a subset of the modes could be taken to represent phononic DOFs of the semiconductor or an implicit solvent. Through \cref{eq:RE_U_ij,eq:RE_V_ij}, the vibrational coordinates can modulate the molecular orbital energies, their couplings, and the electronic correlation between them, but do not contribute to the dye-semiconductor coupling $H_{\rm{int}}$, i.e., the representative model assumes constant non-uniform couplings $\hat W_{ij}(\boldsymbol{Q}) = W_{ij}$.

\subsubsection*{Initial states}
To describe the initial state of the dye-semiconductor system, \cref{eq:initial_state} is used, where $\ket{\Psi(0)}_{\rm{mol}}$ describes the initially photo-excited of the dye, which in the Frank-Condon approximation is given by
\begin{align}
\ket{\Psi(0)}_{\rm{mol}} = \hat \mu \ket{\Phi} \otimes \prod_{\kappa = 0}^{M-1} \ket{\chi}_{\kappa},
\end{align}
where $\hat \mu$ is the electronic dipole operator and $\ket{\Phi}$ represents the electronic ground state in the $N_{\rm{mol}}$ active spin orbitals, e.g., the $N_{\rm{mol}}=4$ case could utilize the $\ket{1100}$ state, describing the doubly occupied HOMO and empty LUMO, or correlated $(2\text{e}, 2\text{o})$ ground state of the embedded HOMO and LUMO space. At zero temperature, the nuclear DOFs $\ket{\chi}_{\kappa}$ would be prepared as harmonic oscillator ground states, and the metal would be initialized to the computational basis state filled to the Fermi level.

\subsubsection*{Observables}
In studying the photoinduced charge injection dynamics, the observable of critical interest is the electronic populations of the dye, obtained through \cref{eq:electronic_populations}. The overall rate of charge transfer is obtained through the time-resolved electronic population of the dye using \cref{eq:electronic_population_spinful} and a fitting of the time series to exponential (or multi-exponential) decay models, such as
\begin{align}
P_{\rm{el}}(t) = \sum_{\eta} A_{\eta} e^{- k_{\eta} t},
\end{align}
where $A_\eta$ and $k_\eta$ are weights and effective rate constants. Beyond the extraction of overall rates, the individual time-resolved spin-orbital occupations $P_i(t)$ provide an orbital-resolved picture of the charge-transfer dynamics, identifying specific molecular orbitals that mediate injection, recombination, and competing pathways. This information would provide a connection between electronic structure and dynamical performance which could be leveraged to guide molecular or interfacial design strategies for dye-sensitized solar cells.

\subsection{Charge transport across molecular junction}

Quantum dynamical simulations addressing the challenging case of an extended Anderson-Newns Hamiltonian including two charged electronic states and two nuclear DOFs have demonstrated that multistate and multimode effects can significantly alter the bond rupture dynamics across various voltage biases and coupling regimes \cite{ke2023current}. Motivated by this, we consider a further extended representative GAN Hamiltonian to describe current-induced bond rupture in a molecular junction for resource estimation of the quantum algorithm. In this example, we set $N_{\rm{mol}}=2$ and resolve the molecular electronic DOFs in the basis of neutral and charged diabatic states. In this representation, the fermionic modes of the molecule in the $\ket{00}$ state encode the neutral ground state of the molecule, states $\hat a_0^\dagger \ket{00}$ and $\hat a_1^\dagger \ket{00}$ encode singly charged anionic states, and $\hat a_0^\dagger \hat a_1^\dagger \ket{00}$ encodes a doubly charged anionic state. Then, $\hat H_{\rm{mol}}$ for the molecular bridge residing between two metallic leads is expressed as 
\begin{align}
\hat H_{\rm{mol}} = \sum_{\kappa = 0 }^M  \frac{\hat P_{\kappa}^2}{2 m_{\kappa}} + \hat U_0(\boldsymbol{\vec{Q}}) + \sum_{i,j=0}^1 \hat U_{ij} (\boldsymbol{\vec{Q}}) \hat a_i^\dagger \hat a_j + \hat V_{01}(\boldsymbol{\vec{Q}}) \hat a_0^\dagger \hat a_0 \hat a_1^\dagger \hat a_1.
\end{align}
Potential $\hat U_0(\boldsymbol{\vec{Q}})$ describes the neutral ground state potential energy surface, $\hat U_{01}(\boldsymbol{\vec{Q}})$ describes the vibronic coupling between the two singly charged diabatic states, and the remainder of the nuclear functions may be defined in terms of the neutral and charged state potential energy surfaces as: 
\begin{align}
\hat U_{00} (\boldsymbol{\vec{Q}}) & = \hat E_{(-1, 0)}(\boldsymbol{\vec{Q}}) - \hat U_0(\boldsymbol{\vec{Q}}) \\ 
\hat U_{11} (\boldsymbol{\vec{Q}}) & = \hat E_{(-1, 1)}(\boldsymbol{\vec{Q}}) - \hat U_0(\boldsymbol{\vec{Q}}) \\ 
\hat V_{01} (\boldsymbol{\vec{Q}}) & = \hat E_{(-2)}(\boldsymbol{\vec{Q}}) - \sum_{i \in \{ 0,1 \}} \hat E_{(-1, i)}(\boldsymbol{\vec{Q}}) + \hat U_0(\boldsymbol{\vec{Q}}),
\end{align}
where $\hat E_{(-1, 0)} (\boldsymbol{\vec{Q}})$ and $\hat E_{(-1, 1)} (\boldsymbol{\vec{Q}})$ denote the singly charged anionic potentials and $\hat E_{(-2)} (\boldsymbol{\vec{Q}})$ is the doubly charged anionic potential. The metallic leads to the left ($L$) and right ($R$) of the molecule are given their own independent electron reservoirs $\mathcal{B}_L$ and $\mathcal{B}_{R}$, 
\begin{align}
\hat H_{\rm{metal}} = \sum_{i \in \mathcal{B}_L \bigcup \mathcal{B}_R} \epsilon_i \hat a_i^\dagger \hat a_i,
\end{align}
and coupling to both leads is parameterized by the coordinate $Q_{M-1}$, 
\begin{align}
\hat H_{\rm{int}} = \sum_{\substack{i \in \mathcal{M} \\ j \in \mathcal{B_{L}} \bigcup \mathcal{B}_R }} \hat W_{ij}(Q_{M-1}) \left( \hat a_i^\dagger \hat a_j + \hat a_j^\dagger \hat a_i \right).
\end{align}
Here, $Q_{M-1}$ need not correspond to a molecule-lead distance, but could also represent an additional reactive coordinate that the molecular conductance is strongly dependent on, such as the dissociation of a side group which significantly alters the $\pi$ conjugation of the molecular bridge  \cite{ke2021unraveling}. We further assume that there are two other internal reactive modes, which contribute to the potentials and couplings in $\hat H_{\rm{mol}}$, but do not vary the molecular conductance, and hence do not contribute to $\hat H_{\rm{int}}$. Finally, we consider up to $10$ non-reactive internal vibrational modes of the bridge. 

\subsubsection*{Initial states}

To simulate the current-induced bond rupture dynamics, one can consider an initial state \cref{eq:initial_state} in which the molecular bridge has its electronic and nuclear DOFs in the ground state, i.e., 
\begin{align}
\ket{\Psi(0)}_{\rm{mol}} = \ket{00} \otimes \prod_{\kappa=0}^{M-1} \ket{\chi}_\kappa,
\end{align}
where $\ket{\chi}_\kappa$ represent the ground vibrational eigenstates in the electronic ground state, e.g., harmonic oscillator and Morse oscillator ground states. A voltage-bias is supplied via the difference in chemical potentials of the metallic leads, $(\mu_L - \mu_R)/e$, where $e$ is the electric charge. With both leads thermally equilibrated at temperature $T = T_L = T_R$, the initial state of the metal is described by
\begin{align}
\hat \rho_{\rm{metal}}(0) = \hat \rho_L (0) \otimes \hat \rho_R (0) = \frac{1}{\mathcal{Z}_L \mathcal{Z}_R}  \bigotimes_{\alpha \in \{ L,R \}}\exp\left(-\sum_{i \in \mathcal{B}_\alpha} \frac{(\epsilon_{i} -\mu_{\alpha})}{k_BT}\, \hat a_{i}^\dagger \hat a_{i}\right).
\end{align}

\subsubsection*{Observables}
To probe the current-induced bond rupture dynamics, the two primary observables of interest are the electronic current and the positions of the reactive nuclear coordinates. Crucially, access to their simultaneous time evolution provides insight that cannot be obtained from either observable in isolation, as correlated changes in molecular geometry and conductance reveal feedback mechanisms between nuclear motion and electronic transport \cite{erpenbeck2023electrical}. The electronic current passing through the molecule is expressed as 
\begin{align}
I(t) = \frac{I_L(t) - I_R(t)}{2},
\end{align}
where $I_{\alpha}(t)$ is the flux of charge from the $\alpha^{\rm{th}}$ lead to the molecule,
\begin{align} \label{eq:charge_flux}
I_{\alpha}(t) = - e \frac{d}{dt} \text{Tr} \left( \hat \rho(t) \hat N_{\rm{metal}, \alpha}^{(\rm{el})} \right), 
\end{align}
where $\hat N_{\rm{metal}, \alpha}^{(\rm{el})} = \sum_{i \in \mathcal{B}_{\alpha}} \hat a_i^\dagger \hat a_i$ is the electron number operator for bath $\alpha \in \{ L, R \}$ \cite{ke2021unraveling}. To estimate \cref{eq:charge_flux}, one can utilize a central finite difference and target
\begin{align}
I_{\alpha}(t) =  -\frac{e}{\Delta t} \left[\text{Tr} \left( \hat \rho\left(t + \frac{\Delta t}{2}\right) \hat N_{\rm{metal}, \alpha}^{(\rm{el})} \right) -  \text{Tr} \left( \hat \rho\left(t - \frac{\Delta t}{2}\right) \hat N_{\rm{metal}, \alpha}^{(\rm{el})} \right) \right]  + \mathcal{O}(\Delta t^2),
\end{align}
where the required expectation values of $\hat N_{\rm{metal}, \alpha}^{(\rm{el})}$ are simply obtained by sampling the $\alpha^{\rm{th}}$ metal register in the computational basis. The position of the reactive nuclear coordinates are obtained as \cref{eq:avg_pos}, from which the stability of these reactive modes under sustained voltage can be assessed over time, and rates of dissociation can be obtained by tracking the reactive coordinates population within a dissociated subspace similarly to \cref{eq:ads_projector}. Additionally, the transient electron population of the molecular bridge, evaluated via \cref{eq:electronic_populations}, provides mechanistic insight into current-induced bond rupture, as occupation of strongly antibonding electronic states has been identified as a key rupture pathway \cite{erpenbeck2020current}. Collectively, access to these electronic and nuclear observables through quantum simulation of classically intractable models would provide a quantitative route to understanding and ultimately controlling molecular junction stability under current flow, with implications for the rational design of more robust and functional molecular electronic devices.

\end{appendix}

    

    


    

    
    

\end{document}